\documentclass[prd,aps,preprintnumbers,floats,floatfix,superscriptaddress,preprintnumbers,
showpacs,eqsecnum,nofootinbib,twocolumn]{revtex4-1}
\usepackage[utf8]{inputenc}
\usepackage{latexsym,array,theorem,mathrsfs,bm,float}
\usepackage{psfrag}
\usepackage{amsfonts,amsmath,amssymb,latexsym,array,afterpage,theorem,color,graphicx,mathrsfs,bm,float,tabularx,here,
,enumerate,multirow,comment}
\usepackage{braket}
\newcommand{\eqnal}[1]{\begin{align}#1\end{align}}
\newcommand{\eqnals}[1]{\begin{align*}#1\end{align*}}
\newcommand{\f}[2]{\frac{#1}{#2}}
\renewcommand{\(}{\left(}
\renewcommand{\)}{\right)}
\renewcommand{\[}{\left[}
\renewcommand{\]}{\right]}
\newcommand{\pd}{\partial}
\renewcommand{\a}{\alpha}

\renewcommand{\b}{\beta}

\renewcommand{\d}{\delta}

\newcommand{\e}{\epsilon}

\newcommand{\ka}{\kappa}

\renewcommand{\l}{\lambda}

\renewcommand{\O}{\Omega}
\newcommand{\si}{\sigma}

\renewcommand{\t}{\tau}

\renewcommand{\th}{\theta}

\renewcommand{\r}{\rho}

\begin{document}
\baselineskip=12pt

\preprint{OCU-PHYS-544, AP-GR-171, NITEP 112, RUP-21-11}

%%%%%%%%%%%%%%%%%%%%%%%
%<<<<<<<<<<<<< TITLE >>>>>>>>>>>>>>>%
%%%%%%%%%%%%%%%%%%%%%%%
\title{
Robustness of particle creation in the formation of a compact object
}
%%%%%%%%%%%%%%%%%%%%%%%
%<<<<<<<<<<< AUTHOR >>>>>>>>>>>>>%
%%%%%%%%%%%%%%%%%%%%%%%
\author{Kazumasa Okabayashi}
\email{okabayashi``at"ka.osaka-cu.ac.jp}
\affiliation{
Department of Mathematics and Physics, Osaka City University, Sumiyoshi,  Osaka 558-8585, Japan}
\author{Tomohiro Harada}
\email{harada``at''rikkyo.ac.jp}
\affiliation{Department of Physics, Rikkyo University, Toshima, Tokyo 171-8501, Japan}
\author{Ken-ichi Nakao}
\email{knakao``at''sci.osaka-cu.ac.jp}
\affiliation{
Department of Mathematics and Physics, Osaka City University, Sumiyoshi,  Osaka 558-8585, Japan}
\affiliation{
Nambu Yoichiro Institute of Theoretical and Experimental Physics, Osaka City University, Sumiyoshi, Osaka City 558-8585, Japan}

%<<<<<<<<<<<<< DATE >>>>>>>>>>>>>>>%
\date{\today}
%%%%%%%%%%%%%%%%%%%%%%%%
%<<<<<<<<<<< ABSTRACT >>>>>>>>>>>>>%
%%%%%%%%%%%%%%%%%%%%%%%%
\begin{abstract}
Hawking predicted that the formation of a black hole by gravitational collapse causes quantum particle creation and the spectrum of the particles is almost thermal.
This phenomenon is called the Hawking radiation. Recently, it has been predicted that the particle creation may drastically change from the Hawking radiation to a strong double burst if the gravitational collapse suddenly stops just before the formation of the event horizon. In contrast with the Hawking radiation, the burst may be so strong that it can be of observational interest even for collapsing objects with astrophysical mass scales.
However, the burst phenomenon has been predicted through the studies of idealized models in which a spherical hollow shell begins to collapse, but stops shrinking and eventually settles down to a static ``star".
Therefore, one might guess that it could be particular
to the hollow shell model. In this paper, we study the particle creation due to the gravitational collapse of a spherical object whose interior is filled with matter. In this model, we obtain similar results to those in the case of the hollow shell model.
This implies that the double burst is a robust property of particle creation by the sudden braking of gravitational collapse.
\end{abstract}

\maketitle

%\tableofcontents
%\setcounter{page}{1}
%\newpage

%%%%%%%%%%%%%%%%%%%%%%%%%%%%%%%%%%%%%%%%%%%%%%%%%%%%%%%%%%%%%%%%%%%%%%%%%
\section{INTRODUCTION}
\label{introduction}
%%%%%%%%%%%%%%%%%%%%%%%%%%%%%%%%%%%%%%%%%%%%%%%%%%%%%%%%%%%%%%%%%%%%%%%%%
The initial vacuum state could evolve to an excited state in curved spacetime.
For example, when an object collapses and a black hole forms, the initial vacuum state changes and eventually settles down to a state in which particles with a thermal spectrum are radiated steadily.
This phenomenon is called Hawking radiation \cite{hawking1974,hawking1975}.
Particle creation is a fundamental phenomenon in quantum field theory in curved spacetime.

In the derivation of the Hawking radiation, the existence of a black hole was originally assumed to justify the geometrical optic approximation.
However, the existence of the black hole is not a necessary condition for the creation of particles with a thermal spectrum.
In fact, particle creation with a thermal distribution can be obtained
when an adiabatic condition is satisfied
even if a black hole does not exist; i.e., one can obtain particle creation with a thermal distribution in a horizonless spacetime
\cite{Barcelo2011}.
This fact means that we cannot know whether an astrophysical object is a black hole even when particle creation with thermal distribution is observed.

A black hole is defined as a complement of the causal past of future null infinity,
and we cannot directly prove from observation that an object is actually a black hole.
Hence, all we can do is to show that an astrophysical object is not a black hole.
As astrophysical objects that cannot be distinguished from black holes,
various alternatives to black holes have been proposed
(see, e.g., Ref. \cite{cardoso2019testing} for a review).

When the collapsing hollow shell stops shrinking just before the formation of an event horizon and becomes an ultra-compact object, it has been found that a couple of bursts in the radiation power of particle creation occur
\cite{PhysRevD.99.044039,Barcel__2019,PhysRevD.100.084028}.
These bursts might be so large that we can observe them although it is very difficult to observe the Hawking radiation for astrophysical objects.
Therefore, it is, in principle, possible to obtain evidence that an object is not a black hole by evaluating the radiation power.
However, the bursts of particle creation might be particular to the hollow shell model whose inside metric is the Minkowskian.
Thus, in this paper, we will consider the gravitational collapse of a spherically symmetric object with a homogeneous density distribution, and evaluate the radiation power when the object eventually becomes ultra compact.

The organization of this paper is as follows.
In Sect. \ref{particle_creation},
we briefly review particle creation from a compact object with spherical symmetry.
A description of how an ultra-compact object forms after the gravitational collapse is given in Sect. \ref{collapse}.
In Sect. \ref{FLRW},
we evaluate the radiation power of the particle creation for the present model and show that a similar result to the case with the hollow shell model is obtained.
In Sect. \ref{mass}, we explicitly show that the distribution of mass of our model is totally different from that of the hollow shell model.
Finally, Sect. \ref{concluding_remarks} is devoted to concluding remarks.
Throughout this paper, we use the units of $c=G=\hbar=1$ in which the speed of light $c$, Newton's gravitational constant $G$, and the Dirac constant $\hbar$ are unity.

%%%%%%%%%%%%%%%%%%%%%%%%%%%%%%%%%%%%%%%%%%%%%%%%%%%%%%%%%%%%%%%%%%%%%%%%%
\section{Particle creation from a compact object with spherical symmetry}
\label{particle_creation}
%%%%%%%%%%%%%%%%%%%%%%%%%%%%%%%%%%%%%%%%%%%%%%%%%%%%%%%%%%%%%%%%%%%%%%%%%
A spacetime is assumed to have spherical symmetry and the asymptotically flat region.
The line element is given as
\eqnal{
	ds^2=-N(t,r_*)(dt^2-dr_*^2)+R(t,r_*)^2d\O^2,
}
where $d\O^2$ is the round metric.
We introduce the null coordinates defined as
$u=t-r_*$, $v=t+r_*$.
By this assumption, $N \rightarrow 1$ and $R \rightarrow r_*$
for  $v \rightarrow \infty$ or $u \rightarrow -\infty$.
An ingoing radial null $v=v_{\rm in}$ turns to an outgoing radial null $u=u_{\rm out}$ after it crosses a point $R=0$.
Since $r_*$ does not necessarily vanish at a point $R=0$ in the curved spacetime, the relation between $v_{\rm in}$ and $u_{\rm out}$ is non-trivial; the relation is denoted by $v_{\text{in}}=G(u_{\text{out}})$.

We define the following quantity;
\eqnal{
	\ka(u_{\text{out}}) :=
	-\f{d}{du_{\text{out}}} \ln \f{dv_{\text{in}} } { du_{\text{out}} }
	=-(\ln G')'(u_{\text{out}}),
}
where the prime denotes the derivative of $u_{\rm out}$.
By the semiclassical approximation, if the adiabatic condition $|\ka'(u_{\rm out})| \ll \ka^2(u_{\rm out})$ is satisfied, the radiation of the massless scalar field with Planck distribution is observed at $u=u_{\rm out}$ in asymptotically flat region \cite{Barcelo2011}.
The temperature of this radiation is estimated as
\eqnal{
	k_B T(u_{\rm out})=\f{\ka(u_{\rm out})}{2\pi},
}
where $k_B$ is Boltzmann constant.
The radiation power $P$ is the total energy per unit time across a sphere.
This represents the integration of the expectation value of the energy-momentum tensor on the sphere.
The energy-momentum tensor of the scalar field is a quadratic operator and we need to regularize it.
By using the point-splitting regularization and the $s$-wave approximation, the radiation power in the asymptotic region is evaluated as \cite{1978_Ford_Parker,birrell_davies_1982}
\eqnal{
	P = \oint \bra{0} T_{~~t}^{r_*} \ket{0} R^2 d\O
	\simeq\f{1}{48\pi} (\ka^2+2\d \ka'),
	\label{radiation_power}
}
where $\d$ respectively takes 1 and $0$ for the minimally and the conformally coupled massless scalars and the vacuum $\ket{0}$ is defined at early time.

For technical simplicity,
we consider a spherically symmetric object
with the homogeneous density distribution.
The geometry of its inside is described by the Friedmann-Lema\^{i}tre-Robertson-Walker(FLRW) metric:
\eqnal{
	ds_{\text{in}}^2 &= - d\t^2 + a^2 (d\chi^2+ \sin^2 \chi d \O^2).
}
The object is assumed to be composed of a perfect fluid with the equation of state $p=w\r$, where $\r$ and $p$ are the energy density and the pressure, respectively.
Here, note that $w$ may depend on $\t$.
Then, the Einstein equations lead to
\eqnal{
 &\dot{a}^2=\f{8\pi}{3}\rho a^2-1 ,
 \label{friedmann}
 \\
 &\dot{\rho}+3H(1+w)\rho=0,
 \label{energy}
}
where a dot means a derivative with respect to $\t$.
We introduce the null coordinates as $V=\eta + \chi$ and $U=\eta - \chi$
where $d\eta = a^{-1}d\t$.

The outside of the object is assumed to be vacuum and
by Birkhoff's theorem is described by the Schwarzschild geometry:
\eqnal{
	ds_{\text{ext}}^2&=-f(r) dt^2 + f(r)^{-1}dr^2+ r^2 d \O^2,
}
where $f(r)=1-2M/r$.
We introduce the null coordinates as $v=t + r_{*}$ and $u=t - r_{*}$
where $dr_{*} = f(r)^{-1}dr$.

\begin{figure}[htbp]
	\includegraphics[width=4cm,keepaspectratio]
	{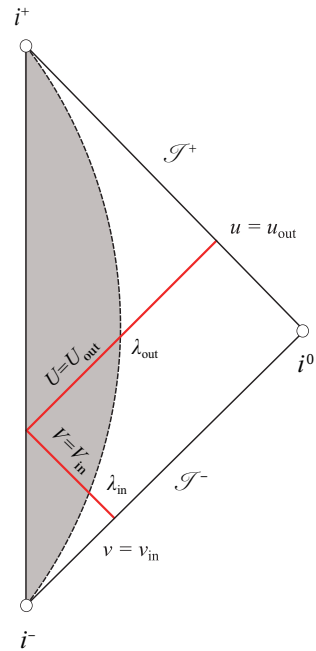}
	\caption{
	The conformal diagram of a collapse to a static compact object.
	The surface of the object moves along a dashed curve whose parameter is $\lambda$, and the matter of the object is represented by a gray region.
	An ingoing radial null at $v=v_{\rm in}$ intersects the surface of the object at $\l=\l_{\rm in}$. After the null crosses a point $R=0$, it finally becomes to an outgoing radial null $u=u_{\rm out}$.
	}
	\label{transmissive_boundary_condition}
\end{figure}

Further, we assume that the surface of the object is comoving with the perfect fluid.
In this setting, the pressure gradient will normally diverge at the surface of the object, and there will be a singular thin shell at the surface.
If a static compact object is formed after a gravitational collapse and the spacetime does not have a horizon, then the conformal diagram of the collapse to a compact object can be schematically depicted in Fig. \ref{transmissive_boundary_condition}.
%%%%%%%%%%%%%%%%%%%%%%%%%
%%%%%%%%%%%%%%%%%%%%%%%%%
%%%%%%%%%%%%%%%%%%%%%%%%%
%%%%%%%%%%%%%%%%%%%%%%%%%
The surface of the object is parametrized by $\l$. The subscript s is used to represent the coordinates of the surface (e.g. $r=r_{\rm s}(\l)$).
An ingoing radial null which intersects the surface of the object at $\l=\l_{\rm in}$ is  $v=v_{\rm s}(\l_{\rm in})$ outside the object whereas $V=V_{\rm s}(\l_{\rm in})$ inside the object.
At the center, the ingoing radial null $V=V_{\rm s}(\l_{\rm in})$ turns to the outgoing radial null $U=U_{\rm s}(\l_{\rm out})$ which intersects the surface of the object at $\l=\l_{\rm out}$.
Since $V_{\rm s}(\l_{\rm in})=U_{\rm s}(\l_{\rm out})$ should hold, we have $\dot{V}_{\rm s}(\l_{\rm in}) d\l_{\rm in}=\dot{U}_{\rm s}(\l_{\rm out}) d\l_{\rm out}$, or equivalently, $d\l_{\rm in}/d\l_{\rm out}=\dot{U}_{\rm s}/\dot{V}_{\rm s}$, where the dot means the derivative with respect to $\l$.
This outgoing null is written as $u=u_{\rm s}(\l_{\rm out})$ outside the object.
Since $v_{\rm in}=v_{\rm s}(\l_{\rm in})$ and $U_{\rm out}=U_{\rm s}(\l_{\rm out})$,
we obtain
\eqnal{
	G'(u_\text{out})
	&=\f{dv_{\rm in}}{du_{\rm out}}
	=\f{\dot{v}_{\rm s}(\l_{\rm in}) d\l_{\rm in}}{\dot{u}_{\rm s}(\l_{\rm out}) d\l_{\rm out}}
	\notag
	\\
	&=
	\f{\dot{v}_{\rm s}(\l_{\rm in})}{\dot{u}_{\rm s}(\l_{\rm out})}
	\f{\dot{U}_{\rm s}(\l_{\rm out})}{\dot{V}_{\rm s}(\l_{\rm in})}
	\notag
	\\
	&=\f{ A(\l_{\rm out} )} { B(\l_{\rm in}) },
	\label{gprime}
}
where
\eqnal{
	A(\l)=\f{\dot{U}_{\text{s}}(\l)}{\dot{u}_{\text{s}}(\l)}
	\quad
	{\rm and}
	\quad
	B(\l)=\f{\dot{V}_{\text{s}}(\l)}{ \dot{v}_{\text{s}}(\l) }.
}
Then, $\ka$ is written as
\eqnal{
	\ka(u_{\text{out}})
	&=
	C(\l_{\text{out}})-\f{A(\l_{\text{out}}) }{B(\l_{\text{in}}) } D(\l_{\text{in} }),
	\label{transmissive_ka}
}
where we introduce
\eqnals{
	C(\l)=
	-\f{1}{\dot{u}_{\rm s}}
	\f{d \ln {A}(\l)} {d\l}
	\quad
	{\rm and}
	\quad
	D(\l)=
	-\f{1}{\dot{v}_{\rm s}} \f{d \ln {B}(\l) }{d\l}.
}
%%%%%%%%%%%%%%%%%%%%%%%%%%%%%%%%%%%%%%%%%%%%%%%%%%%%%%%%%%%%%%%%%%%%%%%%%
\section{Collapse to an ultra-compact object}
\label{collapse}
%%%%%%%%%%%%%%%%%%%%%%%%%%%%%%%%%%%%%%%%%%%%%%%%%%%%%%%%%%%%%%%%%%%%%%%%%
We consider the case in which a static ultra-compact object whose radius is slightly larger than its Schwarzschild radius is a final state of the collapsing object.
The dynamical evolution of the object is classified into several phases specified as intervals with respect to the retarded time $u$ outside the object.
Each phase is labeled by a number $i$.
A radial null entering the object in phase $j$ and escaping from it in phase $i$ is called the null $(i, j)$.

We assume that the surface of the object moves along a timelike curve with $\chi := \chi_0 = {\rm constant}$.
The time coordinate $\t$ inside the object is chosen as the parameter $\l$.
Since $\t$ is the proper time of the surface, we have
\eqnals{
	f\dot{t}_{\rm s}^2-\f{\dot{r}_{\rm s}^2}{f}=1,
}
and hence,
\eqnals{
	\dot{t}_{\rm s}=\f{1}{f}\sqrt{\dot{r}_{\rm s}^2+f
	}.
}

By using this equation, we obtain
\eqnal{
	\dot{u}_{\rm s}&=\dot{t}_{\rm s}-\f{\dot{r}_{\rm s}}{f}=\f{1}{f}\( \sqrt{f + \dot{r}_{\rm s}^2 } - \dot{r}_{\rm s} \),
	\notag
	\\
	\dot{v}_{\rm s}&=\dot{t}_{\rm s}+\f{\dot{r}_{\rm s}}{f}=\f{1}{f}\( \sqrt{f + \dot{r}_{\rm s}^2 } + \dot{r}_{\rm s} \).
	\label{timelike_dot_outside}
}
We also have
\eqnal{
	&\dot{U}_{\rm s}=\dot{V}_{\rm s}= \f{1}{a(\t_{\rm s})},
	\quad
	r_{\rm s}=a(\t_{\rm s}) \sin \chi_0.
	\label{timelike_dot_inside}
}

If the object is composed of the dust $w=0$, its surface shrinks along a radial timelike geodesic:
\eqnal{
	\dot{r}_{\rm s}^2=E^2-1+\f{2M}{r_{\rm s}},
	\quad
	{\rm and}
	\quad
	\ddot{r}_{\rm s}=-\f{M}{r_{\rm s}^2},
	\label{dust_collapse}
}
where $E$ is a conserved specific energy.
However, as Ref. \cite{PhysRevD.99.044039}, we are interested in the case of the matter content with an exotic equation of state which eventually causes the object to stop its gravitational collapse and settle in a static state with a radius $r_{\rm s}=R_{\rm F}$.
Thus, we assume that the dynamics of the object is divided into the following five phases.
%%%%%%%%%%%%%%%%%%%%%%%%%%
\begin{enumerate}[ \textrm{(}i\textrm{)} ]
%%%%%%%%%%%%%%%%%%%%%%%%%%
\item Phase 0: an early-collapse phase for $u<u_0$ $(r_{\rm s}>4M)$.
In this phase, the shell satisfies
	\eqnals{
	f(r_{\rm s})>\f{1}{2}, \quad \dot{r}_{\rm s}^2<1, \quad |\ddot{r}_{\rm s}|<\f{1}{2M}.
	}

\item Phase 1: a late-collapse phase for $u_0<u<u_1$ $(R_{\rm B}<r_{\rm s}<4M)$.
	In this phase, the shell satisfies
	\eqnals{
	&f(r_{\rm s})<\f{1}{2}, \quad f(r_{\rm s}) < \dot{r}_{\rm s}^2 <1,
	\\
	& |\ddot{r}_{\rm s}|=O\( \f{1}{2M} \).
	}
%%%%%%%%%%%%%%%%%%%%%%%%%%
\item Phase 2: an early-braking phase for $u_1<u \le u_2$ $(R_2 \le r_{\rm s}<R_{\rm B})$.
	In this phase, the shell still satisfies
	\eqnals{
		f(r_{\rm s}) \le \dot{r}_{\rm s}^2 <1.
	}
	The conditions $\dot{r}_{\rm s}^2=f(r_{\rm s})$ and $r_{\rm s}=R_{\rm 2}$ hold at $u=u_2$.
%%%%%%%%%%%%%%%%%%%%%%%%%%
\item Phase 3: a late-braking phase for $u_2<u<u_3$ $(R_{\rm F}<r_{\rm s}<R_2)$.
	In this phase, the shell satisfies
	\eqnals{
		\dot{r}_{\rm s}^2<f(r_{\rm s}).
	}
	The collapse of the object stops, i.e., $\dot{r}_{\rm s}^2=0$ at $u=u_3$.
	The condition $\dot{r}_{\rm s}^2 \ll f(r_{\rm s})$ should hold in the domain sufficiently close to $u=u_{3}$.
%%%%%%%%%%%%%%%%%%%%%%%%%%
\item Phase 4: a final static phase for $u>u_3$.
%%%%%%%%%%%%%%%%%%%%%%%%%%
\end{enumerate}
%%%%%%%%%%%%%%%%%%%%%%%%%%
Note that Phase 0 and Phase 1 describe a collapsing dynamics which includes not only the dust case but also the exotic matter cases.

\begin{figure}[htbp]
	\includegraphics[width=8cm,keepaspectratio]
	{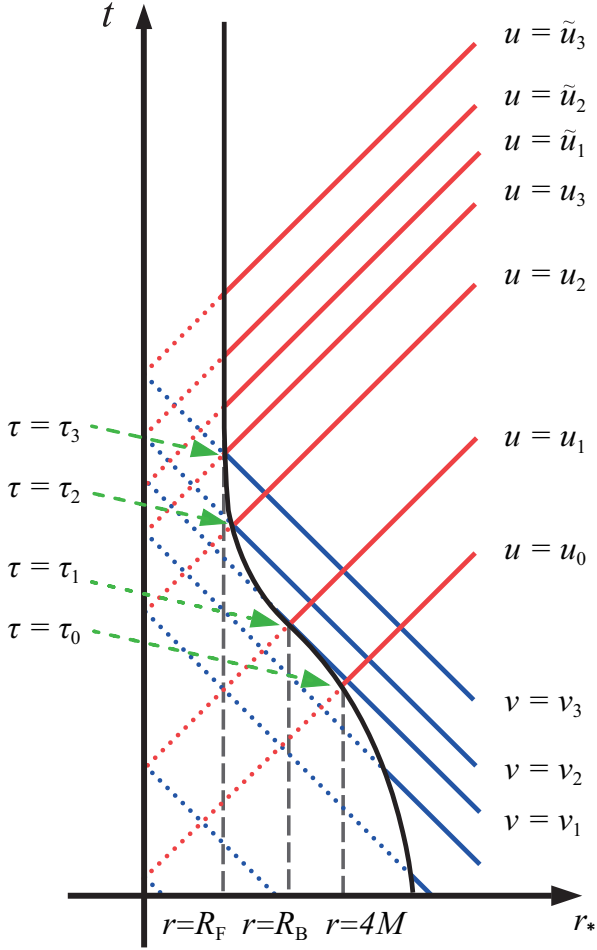}
	\caption{
	The collapse to a compact object is depicted. The black solid curve is the world line of the object's surface.
	The surface is in the collapse phase for $\tau < \tau_1$.
	For $\tau_1<\tau<\tau_3$, it brakes to become static.
	Finally, it is static for $\tau>\tau_3$.
	The vertical axis and the horizontal axis are the time coordinate $t$ and the radial coordinate $r_{*}$ outside the surface, respectively. These coordinates do not cover the region inside the object.
	A blue line represents the ingoing null, and a red line does the outgoing null to which the ingoing null becomes.
	Since these coordinates do not cover the region inside the object, we use dashed lines to specify the relation between the ingoing and the outgoing null curves.
	}
	\label{timelike_shell_sample}
\end{figure}

We  parametrize $R_{\rm F}$ and $R_{\rm B}$ by $\e$ and $\beta$ as
\eqnals{
	R_{\rm F}= 2M(1 + \e^2),
}
and
\eqnals{
R_{\rm B}-R_{\rm F}=2M \e^{2\beta},
}
where $\e^2$ is small and $\beta$ is positive.
This scenario is schematically depicted in Fig. \ref{timelike_shell_sample}.

%%%%%%%%%%%%%%%%%%%%%%%%%
%%%%%%%%%%%%%%%%%%%%%%%%%

%%%%%%%%%%%%%%%%%%%%%%%%%%%%%%%%%%%%%%%%%%%%%%%%%%%%%%%%%%%%%%%%%%%%%%%%%
\section{Evaluation of particle creation}
\label{FLRW}
%%%%%%%%%%%%%%%%%%%%%%%%%%%%%%%%%%%%%%%%%%%%%%%%%%%%%%%%%%%%%%%%%%%%%%%%%
Following the scenario in Sect. \ref{collapse}, we evaluate particle creation through the gravitational collapse of the object.
The functions $A$, $B$, $C$, and $D$ related to $\kappa$ are written as
\eqnal{
	A&=\f{\dot{U}_{\rm s}}{\dot{u}_{\rm s}}=\f{\sin \chi_0}{r_{\rm s}} \(\sqrt{f + \dot{r}_{\rm s}^2}+\dot{r}_{\rm s} \),
	\\
	B&=\f{\dot{V}_{\rm s}}{\dot{v}_{\rm s}}=\f{\sin \chi_0}{r_{\rm s}} \( \sqrt{f + \dot{r}_{\rm s}^2}-\dot{r}_{\rm s} \),
}
\eqnal{
	&
	C:=-\f{1}{\dot{u}_{\rm s}}\f{d}{d\t}\ln A = -a \dot{A}
	\notag
	\\
	&=
	\(-\f{\ddot{r}_{\rm s}}{\sqrt{f+\dot{r}_{\rm s}^2}}+\f{\dot{r}_{\rm s}}{r_{\rm s}} \)
		\( \sqrt{f+\dot{r}_{\rm s}^2} +\dot{r}_{\rm s} \)
	-\f{M\dot{r}_{\rm s}}{r_{\rm s}^2\sqrt{f+\dot{r}_{\rm s}^2}},
	\notag
	\\
	\\
	&
	D:=-\f{1}{\dot{v}_{\rm s}}\f{d}{d\t}\ln B = -a \dot{B}
	\notag
	\\
	&=
	\( \f{\ddot{r}_{\rm s}}{\sqrt{f+\dot{r}_{\rm s}^2}}+\f{\dot{r}_{\rm s}}{r_{\rm s}} \)
		\( \sqrt{f+\dot{r}_{\rm s}^2} -\dot{r}_{\rm s} \)
	-\f{M\dot{r}_{\rm s}}{r_{\rm s}^2\sqrt{f+\dot{r}_{\rm s}^2}}.
	\notag
	\\
}
Hereafter, for notational simplicity,
$r_{\rm s}(\t_{\rm in} )$ and $r_{\rm s}(\t_{\rm out} )$ are denoted by $r_{\rm in}$ and $r_{\rm in}$, respectively.

%%%%%%%%%%%%%%%%%%%%%%%%%%
\begin{enumerate}[ \textrm{(}i\textrm{)} ]
%%%%%%%%%%%%%%%%%%%%%%%%%%
\item Phase 0 ($u<u_0$): the null related to $\kappa$ is classified as $(0,0)$.
%%%%%%%%%%%%%%%%%%%%%%%%%%
In this case, since it is just the beginning of the gravitational collapse, the object is in an almost static state in this phase, and hence the radiation power is so small that we do not discuss it in detail.

%%%%%%%%%%%%%%%%%%%%%%%%%%
\item Phase 1 ($u_0<u<u_1$): the null related to $\kappa$ is classified as $(1,0)$ or $(1,1)$.
%%%%%%%%%%%%%%%%%%%%%%%%%%

We assume that, in this phase, the gravitational collapse proceeds until $|f(r_{\rm s})|$ becomes much less than unity.
In order to evaluate $\ka$, we need to specify the gravitational collapse model.
However, $A(\t_{\rm out})$ is the order of $|f(r_{\rm s})|$ and $C(\t_{\rm out})$ is the order of $M^{-1}$ if $|f(r_{\rm s})|$ is much less than $\dot{r}_{\rm s}^2$.
By contrast, $B(\t_{\rm in})$ and $D(\t_{\rm in})$ are at most the order of $M^{-1}$ for the null (1,0) and (1,1).
Hence, we obtain
\eqnal{
	\ka = \f{1}{4M}
	+O(|f(r_{\rm s})|).
	\label{FLRW_thr}
}
Then, the radiation power is evaluated as
\eqnals{
	P \simeq \f{1}{48\pi}\f{1}{16M^2}.
}
This is the same as the asymptotic value of the radiation power in the formation of the Schwarzschild black hole.
However, in our case, the Schwarzschild black hole does not form as a final product.
In this sense,
we call this type of radiation the transient Hawking radiation.

%%%%%%%%%%%%%%%%%%%%%%%%%%
\item Phase 2 ($u_1<u \le u_2$): the null related to $\kappa$ is classified as $(2,0)$, $(2,1)$, or (2,2).
In this phase, $|\dot{r}_{\rm s}|$ is not necessarily small but at most the order of unity.
%%%%%%%%%%%%%%%%%%%%%%%%%%

As in (ii), for the null (2,0) and (2,1),  $B(\t_{\rm in})$ and $D(\t_{\rm in})$ are at most the order of $M^{-1}$ whereas $A(\t_{\rm out})$ is the order of $|f(r_{\rm s})|$.
By contrast, $\ddot{r}_{\rm out}$ can be too large to ignore terms proportional to $f(r_{\rm s})\ddot{r}_{\rm s}|_{\t=\t_{\rm out}}$ since the gravitational collapse should be stopped within a very short period in the braking phase.
From this, $\kappa$ is roughly evaluated as
\begin{eqnarray}
	\ka
	&=&
	C(\t_{\text{out}}) + O(|f(r_{\rm s})|)
	\notag
	\\
	&\simeq&
	\f{1}{4M}
	-
	\left.
	\f{ f(r_{\rm s})\ddot{r}_{\rm s} }
	{2\dot{r}_{\rm s}^2}
	\right|_{\t_\text{out}},
	\label{kappa_u12}
\end{eqnarray}
where $\dot{r}_{\rm s}^2 \gg f(r_{\rm s})$ has been assumed,
whence we obtain at $u=u_2$
\begin{eqnarray}
	\ka
	&\simeq&
	\f{1}{4\sqrt{2} M} - \f{ 2-\sqrt{2} }{ 2 }  \ddot{r}_{\rm out}.
	\label{kappa_u22}
\end{eqnarray}
Since $|\ka|$ could take a large value for $u_1<u \le u_2$, the radiation power $P$ can be very large in this phase.
We call this radiation the post-Hawking burst.
Note that the first term in Eq.(\ref{kappa_u12}) is the same as $\kappa$ of the transient Hawking radiation, and hence the radiation power also includes the effect of the transient Hawking radiation.
By contrast, the first term in Eq.(\ref{kappa_u22}) is less than $\kappa$ of the transient Hawking radiation.
This is because the collapsing speed is so small that $\dot{r}_{\rm s}^2=f(r_{\rm s}) \ll 1$ holds at $u=u_2$.

For the null (2,2),
$\kappa$ is roughly evaluated as
\begin{eqnarray}
	\ka
	&\simeq&
	\f{1}{4M}
	-
	\[
	\f{\ddot{r}_{\rm out}} {\dot{r}_{\rm out} }
	+
	\f{\ddot{r}_{\rm in} }{ \dot{r}_{\rm in} }
	\]
	\left.
	\f{ f(r_{\rm s}) }
	{2\dot{r}_{\rm s}}
	\right|_{\t_\text{out}}
	,
	\label{kappa_22_u12}
\end{eqnarray}
where $\dot{r}_{\rm s}^2 \gg f(r_{\rm s})$ has been assumed at both $\t=\t_{\rm in}$ and $\t=\t_{\rm out}$,
whence we obtain at $u=u_2$
\begin{eqnarray}
	&&
	\ka
	\simeq
	\f{1}{4\sqrt{2} M}
	\notag
	\\
	&&
	-
	\[
	\f{ 1 }{ \sqrt{2} }
	\f{ \ddot{r}_{\rm out} } { \dot{r}_{\rm out} }
	+\f{ \ddot{r}_{\rm in} } { \dot{r}_{\rm in} }
	\]
	\left.
	\f{f(r_{\rm s})}{(\sqrt{2}+1)\dot{r}_{\rm s}}
	\right|_{\t_\text{out}}
	,
	\label{kappa_22_u22}
\end{eqnarray}
where $\dot{r}_{\rm s}^2 \gg f(r_{\rm s})$ has been assumed only at $\t=\t_{\rm in}$.

%%%%%%%%%%%%%%%%%%%%%%%%%%
\item Phase 3 ($u_2<u<u_3$): the null related to $\kappa$ is classified as $(3,j)$ $(j=0,\cdots,3)$.
For $u_2<u<u_3$, $0<\dot{r}_{\rm s}^2<f(r_{\rm s})$ holds.
At the end of this phase, $\dot{r}_{\rm s}^2$ is much less than $f(r_{\rm s})$.
%%%%%%%%%%%%%%%%%%%%%%%%%%

For (3,0) and (3,1), $\kappa$ is evaluated as
\begin{eqnarray}
	\ka
	&=& C(\t_{\text{out}}) +O(\sqrt{f(r_{\rm s})})
	\simeq
	- \ddot{r}_{\text{out} },
	\label{kappa_u23}
\end{eqnarray}
where $\dot{r}_{\rm s}^2 \ll f(r_{\rm s})$ has been assumed at $\t=\t_{\rm out}$.
For $u_2<u < u_3$, $|\ka|$ could also take a large value since the acceleration of the surface could be large to stop the collapse as for Phase 2, $u_1 < u \le u_2$.

For the null (3,2),
$\kappa$ is roughly evaluated as
\begin{eqnarray}
	&&
	\ka
	\simeq
	- \ddot{r}_{\text{out}}
	-
	\sqrt{f(r_{\rm out})}\f{\ddot{r}_{\rm in}}{|\dot{r}|_{\rm in}},
	\label{kappa_32_u23}
\end{eqnarray}
where $\dot{r}_{\rm s}^2 \gg f(r_{\rm s})$ and $\dot{r}_{\rm s}^2 \ll f(r_{\rm s})$ has been assumed at $\t=\t_{\rm in}$ and $\t=\t_{\rm out}$, respectively.
On the other hand,
\begin{eqnarray}
	&&
	\ka
	\simeq
	- \ddot{r}_{\text{out} }
	-\sqrt{
	\f{f(r_{\rm out} ) } {2 f(r_{\rm in}) }
	}
	\ddot{r}_{\rm in},
	\label{kappa_33_u23}
\end{eqnarray}
where $\dot{r}_{\rm s}^2 = f(r_{\rm s})$ and $\dot{r}_{\rm s}^2 \ll f(r_{\rm s})$ has been assumed at $\t=\t_{\rm in}$ and $\t=\t_{\rm out}$, respectively.

For the null (3,3),
$\kappa$ is roughly evaluated as
\begin{eqnarray}
	&&
	\ka
	\simeq
	- \ddot{r}_{\text{out} }
	- \sqrt{
	\f{ f(r_{\rm out}) } {f(r_{\rm in}) }
	}
	\ddot{r}_{\rm in},
	\label{kappa_33_u23}
\end{eqnarray}
where $\dot{r}_{\rm s}^2 \ll f(r_{\rm s})$ has been assumed at both $\t=\t_{\rm in}$ and $\t=\t_{\rm out}$.

%%%%%%%%%%%%%%%%%%%%%%%%%%
\item Phase 4 ($u_3<u$):
the null related to $\kappa$ is classified as $(4,j)$ $(j=0,\cdots,4)$ and crosses the surface from the inside of the object to its outside in this phase. Hence $C(\t_{\rm out})$ vanishes.
%%%%%%%%%%%%%%%%%%%%%%%%%%

For (4,4), $\kappa$ is exactly zero since the surface is in the static phase.

For (4,0) and (4,1), $A(\t_{\rm out})$ is proportional to $|f(r_{\rm out})|$ whence $B(\t_{\rm in})$ and $D(\t_{\rm in})$ are at most the order of the $M^{-1}$. Hence, $|\kappa|$ is much less than $M^{-1}$.

For (4,2),
$\kappa$ is roughly evaluated as
\eqnal{
	\ka
	&= - A(\t_{\rm out }) \f{ D(\t_{\rm in }) }{ B(\t_{\rm in}) }
	\simeq
	- \e  \f{\ddot{r}_{\rm in} }{| \dot{r}_{\rm in}| },
}
where $\dot{r}_{\rm s}^2 \gg f(r_{\rm s})$ has been assumed at $\t=\t_{\rm in}$.
On the other hand,
when $\dot{r}_{\rm s}^2 = f(r_{\rm s})$ holds at $\t=\t_{\rm in}$,
we obtain
\eqnal{
	\ka
	&= - A(\t_{\rm out }) \f{ D(\t_{\rm in }) }{ B(\t_{\rm in}) }
	\simeq
	- \f{\e}{\sqrt{2}} \f{\ddot{r}_{\rm in} } {|\dot{r}_{\rm in}|}.
}

For (4,3), $\kappa$ is roughly evaluated as
\eqnal{
	\ka
	&= - A(\t_{\rm out }) \f{ D(\t_{\rm in }) }{ B(\t_{\rm in}) }
	\simeq
	- \e  \f{\ddot{r}_{\rm in}} {\sqrt{f(r_{\rm in})} },
}
where $\dot{r}_{\rm s}^2 \ll f(r_{\rm s})$ has been assumed at $\t=\t_{\rm in}$,

Since $\ddot{r}_{\rm in}$ can be very large,
a burst of the particle creation may arise in this phase. We call this the late-time burst.
%%%%%%%%%%%%%%%%%%%%%%%%%%
\end{enumerate}
%%%%%%%%%%%%%%%%%%%%%%%%%%

To evaluate the post-Hawking burst and the late-time burst in detail,
we will consider two concrete braking dynamics in the following,
which are called models A and B.
In model A, the shell slows down exponentially in the braking phase.
In model B,
the shell decelerates by constant acceleration in the braking phase.
In both models A and B, we parametrize $|\dot{r}_{\rm s}|_{\t=\t_1}=\a$ and $R_{\rm B} -R_{\rm F}=2M \e^{2\beta}$ $(\b>0)$, where $R_{\rm B}$ has been defined as the radius of the surface at $\t = \t_1$ when the speed of collapse begins to decrease.
In order to compare the radiation energy of the post-Hawking burst and the late-time burst
with that for the late-collapse phase $(u_0<u<u_1)$, i.e., the transient Hawking radiation,
we evaluate the radiation energy for the collapsing phase:
\eqnal{
	E \simeq P (u_1-u_0)
	\simeq \f{1}{48\pi}\f{\ln \e^{-1}}{2M} \min\{ \beta,1 \},
	\label{E_H}
}
where the time interval $u_1-u_0$ is estimated as $4M \min \{ \beta,1 \} \ln{\e^{-2}}$ in Appendix \ref{time_intervals}.
%%%%%%%%%%%%%%%%%%%%%%%%%%
\subsubsection{Model A: Exponentially slowed-down model}
%%%%%%%%%%%%%%%%%%%%%%%%%%
We assume that the dynamics of the surface is written as $r_{\rm s} - R_{\rm F} \propto e^{-\si \t}$ for $\t_1< \t < \t'_3$ except for a short interval $\t'_3< \t < \t_3$.
For the short interval $\t'_3< \t < \t_3$, the surface of the object is assumed to smoothly settle down to the final static state $R_{\rm F}$.
In this model, we obtain
\eqnal{
	 r_{\rm s}&=
	 2M [1+\e^2+\e^{2\b} e^{-\si (\t-\t_1)}],
	\\
	\si &=
	\f{|\dot{r}_{\rm s}|_{\t=\t_1}}{R_{\rm B} - R_{\rm F}} = \f{\a}{2M \e^{2\beta}}.
}
From $\dot{r}_{\rm s}^2=f(r_{\rm s})$ at $\t=\t_2$,
we obtain a quadratic equation for $X:=e^{-\si(\t_2-\t_1)}$:
\eqnal{
	\a^2 X^2 -\e^{2\beta}X -\e^2=0,
}
where we used $\e^2+\e^{2\b}e^{-\si(\t_2-\t_1)} \ll 1$.
The positive roots of this equation is given as
\eqnal{
	X=\f{ \e^{2\beta} +\sqrt{\e^{4\b}+4\a^2 \e^{2} } } { 2 \a^2 }.
}
If $\e$ is sufficiently small,
$X$ is classified as
\eqnal{
	X
	\simeq
	\begin{cases}
	\dfrac{ \e } { \a }
	& {\rm for } ~~ \beta>1/2
	\\
	\dfrac{1 + \sqrt{1+4\a^2  } }{2\a^2} \e
	& {\rm for } ~~ \beta=1/2
	\\
	\dfrac{ \e^{2\b} } { \a^2 }
	& {\rm for } ~~ 0<\b<1/2
	\end{cases}
	.
	\label{X}
}
In this model, we assume $\t_3-\t_2 \simeq \si^{-1}$.
The radial coordinate of the surface and its derivatives
at $\t=\t_1$ and $\t_2$ are summarized
in Table.\ref{exp_R}.
%%%%%%%%%%%%%%%%%%%%%%%%%%
\begin{table}[H]
\centering
\scalebox{1.2}{
	\begin{tabular}{c|| wc{5em}  wc{5em}  }
		\hline
		$\t$ & $\t_1$ & $\t_2$
		\\ \hline
		$r_{\rm s}(\t)-R_{\rm F}$ & $2 M \e^{2\beta}$ & $2 M \e^{2\beta}X $
		\\ [+.1em]
 		$|\dot{r}_{\rm s}(\t)|$ & $\a$ & $\a X$
 		\\ [+.0em]
		$\ddot{r}_{\rm s}(\t)$ & $\dfrac{\a^2}{2M\e^{2\beta}} $ & $\dfrac{\a^2}{2M\e^{2\beta}}X$
		\\ [+.5em]
		\hline
 	\end{tabular}
	}
\caption{The dynamics of the surface in the braking phase for the exponentially slowed-down model.}
\label{exp_R}
\end{table}
%%%%%%%%%%%%%%%%%%%%%%%%%%
In evaluating the radiation power of particle creation,
it is sufficient to take into account the null of types $(2,0)$, $(2,1)$, $(3,0)$, $(3,1)$ (see Appendix \ref{null_classification}).
In these cases, $C(\t)$ is the dominant term in $\kappa$.
In particular, the following term proportional to $\ddot{r}_{\rm s}$ in $C(\t)$ will be dominant:
\eqnal{
	C_{\rm Acc}
	:=
	-\ddot{r}_{\rm s}
	\(
	1+\f{\dot{r}_{\rm s}}{\sqrt{f+\dot{r}_{\rm s}^2}}
	\).
}
For the early stage in $u_1<u<u_2$, since $\dot{r}^2 \gg f$ holds, the quantity $C_{\rm Acc}$  is estimated as
\eqnal{
	C_{\rm Acc}
	\simeq
	-\f{1 + \e^{2(1-\b)}}{4\a^2 M}.
}
If $\b>1$, we obtain $|\kappa|_{u=u_1} \gg  M^{-1}$ for $0<\e \ll 1$.
As has been mentioned, in order that $|\kappa|$ is larger than
the value of the transient Hawking radiation,
$|\ddot{r}_{\rm s}|$ should be much larger than $M^{-1}$.
Hence, $\b$ should be larger than $1/2$
so that $|\ddot{r}_{\rm s}|_{\t=\t_2} \gg M^{-1}$ for $0<\e \ll 1$.
If $\b>1/2$, $\kappa$ at $u=u_2$ is estimated as
\eqnal{
	\kappa \simeq
	-\f{ 2-\sqrt{2} }{4M\e^{2\beta-1} } \a.
	\label{ka_u2_modelA}
}
For $0<\e \ll 1$, we have $|\kappa|_{u=u_2} \gg |\kappa|_{u=u_1}$.
For the late stage of $u_2<u<u_3$, $|\kappa|$ goes to zero.
The radiation energy for $u_2<u<u_3$ can be roughly evaluated as
\eqnal{
	E
	\simeq \f{1}{2}P(u_3-u_2)
	\simeq \f{1}{48\pi} \f{(2-\sqrt{2})^2 \a }{16M \e^{2\b-1}},
}
where the time interval $u_3-u_2$ is estimated as $2M\a^{-1}\e^{2\b-1}$ is Appendix \ref{time_intervals}.
Comparing it with Eq.(\ref{E_H}), we find that
the radiation energy of the post-Hawking burst is much larger than that of the transient Hawking radiation.

By a similar argument for the late-time burst,
it is sufficient in evaluating $\kappa$ to consider
the null of types $(4, j)$ $(j=0, \cdots, 4)$.
If $\beta$ is larger than $1/2$,
we find that
$\ka$ takes a nearly constant value for $\tilde{u}_1<u<\tilde{u}_2$
\eqnals{
	\ka \simeq
		-\f{\a}{2M\e^{2\beta-1}},
}
and goes to zero for $\tilde{u}_2<u<\tilde{u}_3$.
The radiation energy for $\tilde{u}_1<u<\tilde{u}_2$ is given by
\eqnal{
	E
	\simeq P (\tilde{u}_2-\tilde{u}_1)
	\simeq \f{1}{48\pi}\f{\a \ln (\a \e^{-1}) }{2M \e^{2\b-1}},
}
where the time interval $\tilde{u}_2-\tilde{u}_1$ is estimated as $2M\a^{-1}\e^{2\beta-1} \ln (\a \e^{-1})$ in Appendix \ref{time_intervals}.
Comparing it with Eq.(\ref{E_H}), the radiation energy of the late-time burst is also much larger than that of the transient Hawking radiation.
%%%%%%%%%%%%%%%%%%%%%%%%%%
\subsubsection{Model B: Constant-deceleration model}
%%%%%%%%%%%%%%%%%%%%%%%%%%
We consider that the constant deceleration $\tilde{a}$ of the surface is given by
\begin{eqnarray}
	\ddot{r}_{\rm s}=\tilde{a}= \f{\a^2}{ 2( R_{\rm B} - R_{\rm F} ) },
	\label{deceleration}
\end{eqnarray}
for $\t_1< \t < \t_3$ so that $|\dot{r}_{\rm s}|_{\t=\t_3}$ vanishes.
The deceleration $\tilde{a}$ is much larger than $1/4M$ for $0<\e \ll 1$.
In this model,
the radial coordinate of the surface is
\eqnal{
	r_{\rm s}=
	2M[1+\e^2+ \e^{2\b} (Y-1)^2 ],
}
where
\eqnal{
	Y := \f{\t - \t_1}{\t_3 - \t_1}=\f{\a}{4M\e^{2\b}}(\t-\t_1).
}
In this model, the proper time of the surface, the radial coordinate of the surface and its derivative at $\t=\t_1$, $\t_2$, and $\t_3$ are summarized in Table.\ref{const_R}.
%%%%%%%%%%%%%%%%%%%%%%%%%%
\begin{table}[H]
\centering
\scalebox{1.2}{
	\begin{tabular}{c || wc{5em}  wc{6em}  wc{2em} }
		\hline
		$\t$ & $\t_1$ & $\t_2$ & $\t_3$
		\\ [+.0em] \hline
		$\t_3 - \t$ & $\a \tilde{a}^{-1} $ & $\e \tilde{a}^{-1}$ & $0$
		\\ [+.3em]
		$r_{\rm s}(\t)-R_{\rm F}$ & $2 M \e^{2\beta}$ & $2 M \a^{-2} \e^{2\beta+2}$ & $0$
 		\\ [+.0em]
		$\dot{r}_{\rm s}(\t)$ & $\a$ & $\e$  & $0$
		\\ \hline
 	\end{tabular}
}
\caption{The dynamics of the surface in the braking phase
	for the constant-deceleration model.}
\label{const_R}
\end{table}
%%%%%%%%%%%%%%%%%%%%%%%%%%
From Appendix \ref{null_classification},
it is sufficient in evaluating $\kappa$ in the epoch of the braking phase of this model to consider the null of types $(2,0)$, $(2,1)$, $(3,0)$, $(3,1)$.
In these cases, $C(\t)$ is the dominant term in $\kappa$.
The quantity $C_{\rm Acc}$ is estimated as
\eqnal{
	C_{\rm Acc} =
	-\f{\a^2}{4M\e^{2\b}}
	\[ 1- \f{\a(1-Y)}{\sqrt{\e^2+(\e^{2\b}+\a^2)(1-Y)^2}} \].
}
For the early stage in $u_1<u<u_2$, i.e., the stage with $\dot{r}^2 \gg f$, the quantity $C_{\rm Acc}$  is estimated as
\eqnal{
	C_{\rm Acc} \simeq
	-\f{1+\e^{2\beta-2}(1-Y)^2}{8M\e^{2\b-2}(1-Y)^2}.
}
If $\b>1$, we have $|\kappa| \gg M^{-1}$ for $0<\e \ll 1$.
On the other hand,
we find that $\kappa$ for $u_2 < u < u_3$ take a nearly constant value:
\eqnal{
	\ka \simeq -\f{\a^2}{4M\e^{2\beta}}.
}
For $0<\e \ll 1$,
we have $|\kappa|_{u_2 < u < u_3} \gg |\kappa|_{u=u_1}$.
The radiation energy for $u_2 < u < u_3$ is given as
\eqnal{
	E \simeq P(u_3-u_2) \simeq \f{1}{48\pi} \f{\a^2}{4M \e^{2\b}},
}
where the time interval $u_3-u_2$ is estimated as $4M\a^{-2}\e^{2\b}$ in Appendix \ref{time_intervals}.
Comparing this with Eq.(\ref{E_H}), we find that the radiation energy of the post-Hawking burst is much larger than that for the transient Hawking radiation.

By a similar argument for the late-time burst,
it is sufficient in evaluating $\kappa$ to consider the null of types $(4, j)$ $(j=0,\cdots, 4)$.
If $\b>1/2$ and $0< \e \ll 1$,
$|\ka|$ for $\tilde{u}_1 \le u < \tilde{u}_2$ increases from
$\a / (4M \e^{2\beta-1})$ to $\a^2 / (4M\e^{2\beta})$.
For $\tilde{u}_2 \le u < \tilde{u}_3$, $\ka$ takes a nearly constant value:
\eqnals{
	\ka \simeq -\f{\a^2}{4M\e^{2\beta}}.
}
For $0<\e \ll 1$,
we have $|\kappa|_{\tilde{u}_2 < u < \tilde{u}_3} \gg |\kappa|_{u=\tilde{u}_1}$.
The radiation energy for $\tilde{u}_2 < u < \tilde{u}_3$ is given by
\eqnal{
	E \simeq P(\tilde{u}_3 - \tilde{u}_2) \simeq \f{1}{48\pi}\f{\a^2}{4M\e^{2\beta}},
}
where the time interval $\tilde{u}_3-\tilde{u}_2$ is estimated as $4M\a^{-2}\e^{2\b}$ in Appendix \ref{time_intervals}.
Comparing this with Eq.(\ref{E_H}), we find that the radiation energy of the late-time burst is also much larger than that for the transient Hawking radiation.

%%%%%%%%%%%%%%%%%%%%%%%%%%%%%%%%%%%%%%%%%%%%%%%%%%%%%%%%%%%%%%%%%%%%%%%%%
\section{The Misner-Sharp mass inside the shell and that of the shell}
\label{mass}
%%%%%%%%%%%%%%%%%%%%%%%%%%%%%%%%%%%%%%%%%%%%%%%%%%%%%%%%%%%%%%%%%%%%%%%%%
We obtain the radiation power of the transient Hawking radiation, post-Hawking burst, and late-time burst caused by the gravitational collapse of a homogeneous spherical object with a singular surface, i.e., an infinitesimally thin shell. Their leading terms are similar to those in the hollow shell model, except for the coefficients.
This result seems to be non-trivial.
However, there is a possibility that
the mass except for that of the shell is so small compared to that of the shell that the contribution from the inside matter to the radiation power is negligible, and hence
the radiation power in our model becomes similar to the hollow shell model.
If this is true, our result is not very new because it is physically same as the hollow shell model in such a case.
Hence, we will show that our model is different from the hollow shell model
by comparing the mass inside the shell and that of the shell.
Although, in General Relativity, it is a difficult problem to define a mass, there is a quasi-local definition of a mass in a spherically symmetric spacetime, which is called the Misner-Sherp mass
\cite{PhysRevD.53.1938}.
We adopt this for our present purpose.

Using the double null coordinates, the metric of a spherically symmetric spacetime is given as
\eqnal{
	ds^2=-\tilde{N} (u,v) du dv + \tilde{R}(u,v)^2 d\O^2,
}
where $u$ and $v$ are respectively outgoing and ingoing null coordinates.
The expansion for outgoing and ingoing null congruence can be written as
$\th_{+}=2\tilde{R}^{-1}\pd_{u}\tilde{R}$ and $\th_{-}=2\tilde{R}^{-1} \pd_{v}\tilde{R}$, respectively.
Then, the Misner-Sherp mass $\mathcal{M}$ is defined as
\eqnal{
	\mathcal{M}=\f{\tilde{R}^3}{2} \( \tilde{N}^{-1} \th_+ \th_- + \f{1}{\tilde{R}^2}\).
}
The masses of the object seen from outside and inside the shell are
denoted by $\mathcal{M}_{\pm}$, respectively.
Then, the mass of the shell is given as $\mathcal{M}_{+}-\mathcal{M}_{-}$.

In our model, the metric outside the shell is the Schwarzschild metric and
we obtain
\eqnal{
	\mathcal{M}_{+}=\f{r_{\rm s}}{2}\[1-f(r_{\rm s})\]=M.
	\label{m_plus}
}
By contrast, the metric inside the shell is a closed FLRW metric and
we obtain
\eqnal{
	\mathcal{M}_{-}&=\f{1}{2} a(\t) \[ 1+\dot{a}(\t)^2 \] \sin^3 \chi
		= \f{4 \pi }{3} \r r_{\rm s}^3.
	\label{m_minus}
}
From Eqs. (\ref{m_plus}) and (\ref{m_minus}), the mass of the shell can be evaluated.
To do this, we have to specify the collapsing phase and the braking phase.
As the braking phase, we use the models A and B.
By contrast, as the collapsing phase,
we use the Oppenheimer-Snyder collapse.

In the Oppenheimer-Snyder collapse,
matter inside the object is assumed to be dust\cite{PhysRev.56.455}.
A solution of Eqs.(\ref{friedmann}), (\ref{energy}) is given by
\eqnals{
	&
	a(\eta)=\f{1}{2}a_{\rm max}(1+\cos \eta), \quad
	\t(\eta)=\f{1}{2}a_{\rm max}(\eta+\sin \eta),
}
where we choose the origin of the conformal time $\eta=0$
such that the object starts to collapse, and
$a_{\rm max}$ is the scale factor at $\eta=0$.
Let $\eta_1$ be the conformal time such that the object starts to brake so that $\t(\eta_1)=\t_1$.
At $\eta=\eta_1$, the radius and the velocity of the shell are assumed to be
\eqnal{
	r_{\rm s}(\tau_1)=2M+2M \e^{2\beta},
	\quad
	\left.
	\f{dr_{\rm s}}{d\t}
	\right|_{\tau=\t_1}
	= - \a \quad (\a>0).
	\notag
	\\
}

By contrast, in the stage of the Oppenheimer-Snyder collapse, there is no shell, $\mathcal{M}_{+}-\mathcal{M}_{-}=0$, and we have
\eqnal{
	M=\f{4\pi}{3}\r r_{\rm s}^3 = \f{1}{2}a_{\rm max} \sin^3 \chi_0.
}
After all, we obtain the following equations
\eqnals{
	&
	a_{\rm max} =\f{2M}{\sin^3 \chi_0},
	\\
	\quad
	&
	\eta_1 =2\arccos\[ \sin \chi_0 \sqrt{\f{r_{\rm s}(\tau_1)}{2M}} \],
	\\
	\quad
	&
	\a = \tan \f{\eta_1}{2} \sin \chi_0.
}
When the parameters $\chi_0$ and $\beta$ are fixed,
the remaining parameters, $a_{\rm max}, \eta_1$, and $\a$, are also fixed.
Thus, we could continuously connect
the Oppenheimer-Snyder collapse and the braking phase.

\begin{figure}[htbp]
	\includegraphics[width=8cm,keepaspectratio]
	{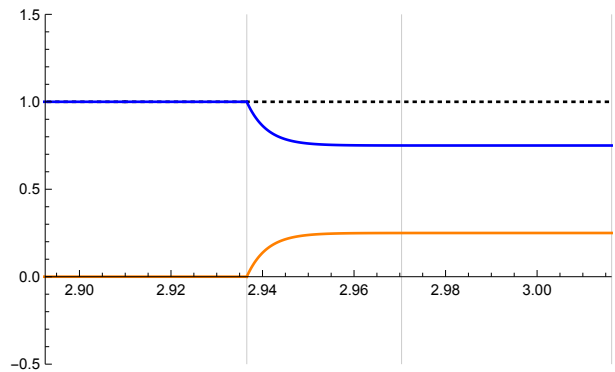}
	\caption{
		The relation between the mass inside the shell and
		that of the shell in the exponentially slowed-down model.
		The blue and orange lines are respectively
		the mass inside the shell and that of the shell.
		The dashed line is the mass outside the shell
		which corresponds to $M$.
		In this figure, we choose $M=1$, $\beta=2/3$, $\e=10^{-2}$, $\chi_0=\pi/3$.
	}
	\label{exp_mass}
\end{figure}

\begin{figure}[htbp]
	\includegraphics[width=8cm,keepaspectratio]
	{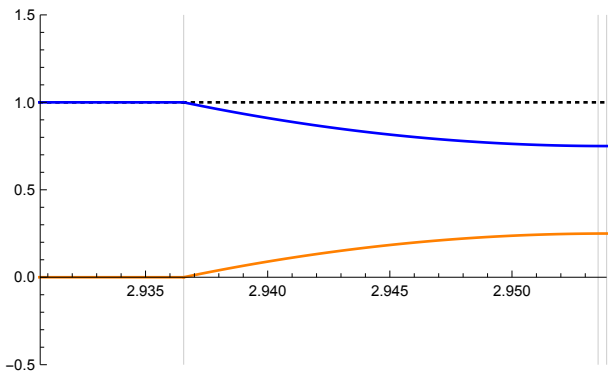}
	\caption{
		The relation between the mass inside the shell and
		that of the shell in the constant-deceleration model.
		The blue and orange lines are respectively
		the mass inside the shell and that of the shell.
		The dashed line is the mass outside the shell
		which corresponds to $M$.
		In this figure, we choose $M=1$, $\beta=2/3$, $\e=10^{-2}$, $\chi_0=\pi/3$.
	}
	\label{acc_mass}
\end{figure}

In Figs. \ref{exp_mass} and \ref{acc_mass},
the relation between the mass inside the shell and that of the shell is given
for models A and B, respectively.
As seen in both figures, the mass of the shell increases through the braking phase although the mass of the shell is zero in the collapsing phase.
However, the mass inside the shell could be larger than the mass of the shell, and hence is not necessarily negligible.
Our result is non-trivial because it is physically different from the hollow shell model.

%%%%%%%%%%%%%%%%%%%%%%%%%
%%%%%%%%%%%%%%%%%%%%%%%%%
%%%%%%%%%%%%%%%%%%%%%%%%%
%%%%%%%%%%%%%%%%%%%%%%%%%
%%%%%%%%%%%%%%%%%%%%%%%%%
%%%%%%%%%%%%%%%%%%%%%%%%%
%%%%%%%%%%%%%%%%%%%%%%%%%
%%%%%%%%%%%%%%%%%%%%%%%%%

%%%%%%%%%%%%%%%%%%%%%%%%%%%%%%%%%%%%%%%%%%%%%%%%%%%%%%%%%%%%%%%%%%%%%%%%%
\section{Concluding remarks}
\label{concluding_remarks}
%%%%%%%%%%%%%%%%%%%%%%%%%%%%%%%%%%%%%%%%%%%%%%%%%%%%%%%%%%%%%%%%%%%%%%%%%

In the current analysis, we treat the spherically symmetric object with the homogeneous density distribution and evaluate the radiation power of the particle creation.

Through the evaluation of the quantity $\kappa$,
we show that the transient Hawking radiation occurs for the collapsing phase,
and that the post-Hawking burst and the late-time burst occur for the braking phase.
The transient Hawking radiation is obtained in the hollow shell model studied in Ref.\cite{PhysRevD.99.044039}, and hence the transient Hawking radiation in the present model is similar to the hollow shell model.
The radiation power for the braking phase depends on the detail of the dynamics of the surface, and hence we consider the two scenarios (model A and B).
For the two examples, we find that these bursts could have the radiation power much larger than the transient Hawking radiation if the acceleration is sufficiently large.
The radiation power of our model is different from that of
the hollow shell model in the numerical factor of the order
of unity (e.g. the parametrization of $|\dot{r}_{\rm s}|_{\t=\t_1}$ and the numerical factor in Eq.(\ref{ka_u2_modelA})).
This does not come from the physical difference
between the models but is rather because the present analysis
is more accurate than the previous one. In spite of this little
difference, it is more important that the exponents of $\epsilon$
for the post-Hawking burst and the late-time burst in the present
matter-filled model completely agree with those in the hollow
shell model.
In this sense, these two bursts we obtained are also similar to the hollow shell model.
This result implies that the bursts of the particle accompanied by the sudden braking of the gravitational collapse has a robust property.
In addition to the radiation power of particle creation,
the radiation energy of the transient Hawking radiation, the post-Hawking burst, and the late-time burst also have the robust property since we find that the time intervals for these phase are also similar to those in the hollow shell model.

We discuss why the similar result is obtained.
In the collapsing phase,
the contribution from matter inside the surface becomes subdominant when the radius of the surface is close to the Schwarzschild radius,
and hence the transient Hawking radiation is obtained.
In the braking phase,
the leading term of the radiation is proportional to the acceleration of the surface
since the absolute value of the acceleration of the shell becomes sufficiently large.
The contribution from matter inside the surface does not affect the exponent of $\e$ in the leading term.
Hence, the post-Hawking burst and the late-time burst in the present model are also similar to the hollow shell case.

By contrast,
one might guess that our model is
effectively the same as the hollow shell model.
To compare our model with the hollow shell model,
we evaluate Misner-Sherp mass inside the shell and that of the shell for the timelike shell model.
Since we find that Misner-Sherp mass inside the shell is dominant in the total mass,
our shell model is different from the hollow shell model.
Nevertheless, we obtain a similar radiation power from our model with the hollow shell model.
This implies that this type of particle creation does not depend on the inside detail and has a robust property.
At this point, this robustness is just implied from our model and not proven in general.
Hence, this robustness should be examined in a more general situation.

%\newpage
\section*{Acknowledgments}
%The authors are grateful to ??? for helpful comments.
This work was supported in part by JSPS KAKENHI Grant Numbers JP21J15676(KO), JP19K03876, JP19H01895, JP20H05853 (TH), and JP21K03557(KN).
~~\\
%%%%%%%%%%%%%%%%%%%%%%%%%%%%%%%%%%%%%%%%%%%%%%%%%%%%%%%%%%%%%%%%%%%%%%%
%\newpage
%%%%%%%%%%%%%%%%%%%%%%%%%%%%%%%%%%%%%%%%%%%%%%%%%%%%%%%%%%%%%%%%%%%%%%%

%%%%%%%%%%%%%%%%%%%%%%%%%%%%%%%%%%%%%%%%%%%%%%%%%%%%%%%%%%%%%%%%%%%%%%%
\appendix
%%%%%%%%%%%%%%%%%%%%%%%%%%%%%%%%%%%%%%%%%%%%%%%%%%%%%%%%%%%%%%%%%%%%%%%%%
%%%%%%%%%%%%%%%%%%%%%%%%%%
\section{Time intervals}
\label{time_intervals}
%%%%%%%%%%%%%%%%%%%%%%%%%%
In order to estimate the total radiation energy, we evaluate a time interval of each phase.
From Eqs.(\ref{timelike_dot_outside}) and (\ref{timelike_dot_inside}),
we obtain
\begin{eqnarray}
u \simeq
	\left\{
	\begin{array}{ll}
		\displaystyle{
		\f{\t}{\sqrt{f(r_{\rm s})}} +\text{const.}
		}
		\quad &(u<u_0)
		\\
		\displaystyle{
		-4M \ln\[ \f{r_{\rm s}}{2M}-1 \]-2r_{\rm s} +\text{const.}
		}
		\quad &(u_0<u<u_2)
		\\
		\displaystyle{
		\displaystyle{
		\int \f{d \t }{ \sqrt{f(r_{\rm s})} } + \text{const.}
		}
		}
		\quad &(u_2<u<u_3)
		\\
		\displaystyle{
		\f{\t}{\e} +\text{const.}
		}
		\quad &(u_3<u)
	\end{array}
	\right.
	\notag
	\\
	\label{FLRW_u_interval}
\end{eqnarray}
From Eq.(\ref{FLRW_u_interval}), we obtain the time interval $u_1-u_0$ as
\eqnal{
u_1-u_0 \simeq
	\left\{
	\begin{array}{ll}
		\displaystyle{
		4M\beta \ln \e^{-2}
		}
		\quad &(0<\beta<1)
		\\
		\displaystyle{
		4M \ln \e^{-2}
		}
		\quad &(\beta \ge 1)
	\end{array}
	\right.
	,
}
where $r_{\rm s}=4M$ at $u=u_0$ and $r_{\rm s}=R_{\rm B}$ at $u=u_1$.
The interval $\tilde{u}_3-u_3$ is given as
\eqnal{
	\tilde{u}_3-u_3 &= \f{2M}{\e \sin \chi_0}(\tilde{U}_3-U_3)
	=\f{4\chi_0 M}{\e \sin \chi_0},
}
where we used $\tilde{U}_3 - U_3 = 2 \chi_0$.
The other intervals depend on the detail of the model in the braking phase.
In what follows, we focus on $\b>1/2$
since the radiation much larger than the transient Hawking radiation is obtained in this case.
%%%%%%%%%%%%%%%%%%%%%%%%%%
\subsubsection{Model A}
%%%%%%%%%%%%%%%%%%%%%%%%%%
From Eq.(\ref{X}) for $\b>1/2$, we obtain
\eqnal{
	\t_2-\t_1=\f{1}{\si} \ln \a \e^{-1},
}
and the time interval $u_2-u_1$ is evaluated through Eq.(\ref{FLRW_u_interval}) as
\eqnal{
	u_2-u_1
	\simeq
	\left\{
	\begin{array}{ll}
	4M (1-\b) \ln \e^{-2} & ( 1/2 < \b <1 ) \\
	4M  & \( \b =1 \) \\
	4M \e^{2\b - 2} & \( \b >1 \)
	\end{array}
	\right.
	.
}
To evaluate the time interval $u_3-u_2$, we have to determine the $\t_3-\t_2$.
If we assume $\t_3-\t_2=\si^{-1}$, the interval $u_3-u_2$ is evaluated through Eq.(\ref{FLRW_u_interval}) as
\eqnal{
	u_3-u_2 \simeq \f{\t_3-\t_2}{\e} \simeq 2M \a^{-1} \e^{2\b -1}.
}
As for the time intervals $\t>\t_3$, the surface of the object is static,
and hence the time interval $\tilde{u}_2-\tilde{u}_1$ is evaluated as
\eqnal{
	\tilde{u}_2-\tilde{u}_1
	&=
	\f{1}{\e} (\tilde{\t}_2 -\tilde{\t}_1)
	\simeq
	\f{1}{\e} (\t_2 -\t_1)
	\notag
	\\
	&\simeq
	2M \a^{-1} \e^{2\b-1} \ln \a \e^{-1},
}
and the same calculation can be done for the time interval $\tilde{u}_3-\tilde{u}_2$:
\eqnal{
	\tilde{u}_3-\tilde{u}_2
	&=
	\f{1}{\e} (\tilde{\t}_3 -\tilde{\t}_2)
	\simeq
	\f{1}{\e} (\t_3 -\t_2)
	\notag
	\\
	&\simeq
	2M \a^{-1} \e^{2\b-1}.
}

%%%%%%%%%%%%%%%%%%%%%%%%%%
\subsubsection{Model B}
%%%%%%%%%%%%%%%%%%%%%%%%%%
From Fig.\ref{const_R}, we obtain
\eqnal{
	\t_2-\t_1 \simeq \a \tilde{a}^{-1},
}
and hence the time interval $u_2-u_1$ is evaluated as
\eqnal{
	u_2-u_1
	\simeq
	\left\{
	\begin{array}{ll}
	4M (1-\b) \ln \e^{-2} & ( 1/2 < \b <1 ) \\
	4M  & \( \b =1 \) \\
	4M \e^{2\b - 2} & \( \b >1 \)
	\end{array}
	\right.
	.
}
The interval $u_3-u_2$ is given as
\eqnal{
	u_3-u_2 \simeq
	\f{\t_3-\t_2}{\e}
	=\tilde{a}^{-1}
	=4M\a^{-2}\e^{2\b}.
}
As for the time intervals $\t>\t_3$, the surface of the object is static,
and hence the time interval $\tilde{u}_2-\tilde{u}_1$ is evaluated as
\eqnal{
	\tilde{u}_2-\tilde{u}_1
	&=
	\f{1}{\e} (\tilde{\t}_2 -\tilde{\t}_1)
	\simeq
	\f{1}{\e} (\t_2 -\t_1)
	\notag
	\\
	&\simeq
	\a \e^{-1} \tilde{a}^{-1}
	=4M\a^{-1}\e^{2\b-1},
}
and the same calculation can be done for the time interval $\tilde{u}_3-\tilde{u}_2$:
\eqnal{
	\tilde{u}_3-\tilde{u}_2
	&=
	\f{1}{\e} (\tilde{\t}_3 -\tilde{\t}_2)
	\simeq
	\f{1}{\e} (\t_3 -\t_2)
	\notag
	\\
	&\simeq
	\tilde{a}^{-1}
	=4M\a^{-2}\e^{2\b}.
}

%%%%%%%%%%%%%%%%%%%%%%%%%%
\section{The null classification of $\kappa$}
\label{null_classification}
%%%%%%%%%%%%%%%%%%%%%%%%%%
In evaluating $\kappa$ for a phase,
we have to know the type of the null $(i,j)$ for the phase.
For $u_0 < u < u_1$, the null is classified as $(1,0)$ or $(1,1)$.
For $u> u_3$, the null is classified as $(4,j)$ $(j=0, \cdots, 4)$.
By contrast, the classification of the null for $u_1< u < u_3$
depends on the detail of the dynamics of the surface.
In general, the null for $u_1< u < u_3$ might be
classified as $(2,j)$ $(j=0, \cdots, 2)$ or $(3,j)$ $(j=0, \cdots, 3)$.
However, we show that
the null classified as $(2,2)$ or $(3,3)$ does not exist in model A and B.
%%%%%%%%%%%%%%%%%%%%%%%%%%
\subsubsection{Model A}
%%%%%%%%%%%%%%%%%%%%%%%%%%
In the braking phase, the conformal time $d\eta := d\t/a$ is written as
\eqnal{
	d\eta \simeq \f{\sin \chi_0}{2M} d\t.
}

For the early-braking phase,
we obtain the interval of the conformal time $\eta_2-\eta_1$ as
\eqnal{
	\eta_2 - \eta_1=\f{\sin \chi_0}{2M}(\t_2 - \t_1).
}
For $\b>1/2$ and $\e \ll 1$, we find from Eq.(\ref{X}) $X := e^{ -\si (\t_2-\t_1) }=\e/\a$,
and hence the interval of the conformal time $\eta_2-\eta_1$ is evaluated as
\eqnal{
	\eta_2 - \eta_1=\f{\e^{2\b}} {\a} \sin \chi_0 \ln \f{\a}{\e} \ll 1.
}
This means that
if any null crosses inwardly the surface for $\t_1<\t<\t_2$,
the time interval of the conformal time $\eta_2-\eta_1$ is too small for the null to cross outwardly the surface for $\t_1<\t<\t_2$.
Hence, we do not have to consider the null $(2,2)$ in evaluation of $\kappa$.

For the late-braking phase, by the similar argument,
the interval of the conformal time $\eta_3-\eta_2$ is given as
\eqnal{
	\eta_3 - \eta_2=\f{\sin \chi_0}{2M}(\t_3 - \t_2).
}
In model A, we choose the interval of the comoving time $\t_3-\t_2=\si^{-1}$,
and hence the interval of the conformal time $\eta_3-\eta_2$ is evaluated as
\eqnal{
	\eta_3 - \eta_2=\f{\e^{2\b}} {\a} \sin \chi_0 \ll 1.
}
This also means that
if any null crosses inwardly the surface for $\t_2<\t<\t_3$,
the interval of the conformal time $\eta_3-\eta_2$ is too small for the null to cross outwardly the surface for $\t_2<\t<\t_3$.
Hence, we do not have to consider the null $(3,3)$ in evaluation of $\kappa$.
%%%%%%%%%%%%%%%%%%%%%%%%%%
\subsubsection{Model B}
%%%%%%%%%%%%%%%%%%%%%%%%%%
For the early-braking phase,
we obtain the interval of the conformal time $\eta_2-\eta_1$ as
\eqnal{
	\eta_2 - \eta_1=\f{\sin \chi_0}{2M}(\t_2 - \t_1).
}
From Table.\ref{const_R}, we obtain the interval of the comoving time $\t_2-\t_1=(\a-\e)\tilde{a}^{-1}$, and hence
the interval of the conformal time $\eta_2-\eta_1$ is evaluated as
\eqnal{
	\eta_2 - \eta_1 \simeq \f{2\e^{2\b}} {\a^2} \sin \chi_0  \ll 1.
}
Hence, if any null crosses inwardly the surface for $\t_1<\t<\t_2$,
the time interval of the conformal time $\eta_2-\eta_1$ is too small for the null to cross outwardly the surface for $\t_1<\t<\t_2$.
Thus, we do not have to consider the null $(2,2)$ in evaluation of $\kappa$.

For the late-braking phase, by the similar argument,
the conformal time $\eta_3-\eta_2$ is given as
\eqnal{
	\eta_3 - \eta_2=\f{\sin \chi_0}{2M}(\t_3 - \t_2).
}
From Table.\ref{const_R}, we obtain the interval of the comoving time $\t_3-\t_2=\e \tilde{a}^{-1}$, and hence
the interval of the conformal time $\eta_3-\eta_2$ is evaluated as
\eqnal{
	\eta_3 - \eta_2=\f{\e^{2\b+1}} {\a} \sin \chi_0 \ll 1.
}
Hence, if any null crosses inwardly the surface for $\t_2<\t<\t_3$,
the conformal time interval $\eta_3-\eta_2$ is too small for the null to cross outwardly the surface for $\t_2<\t<\t_3$.
Thus, we do not have to consider the null $(3,3)$ in evaluation of $\kappa$.

%%%%%%%%%%%%%%%%%%%%%%%%%%
%%%%%%%%%%%%%%%%%%%%%%%%%%
%\newpage
%%%%%%%%%%%%%%%%%%%%%%%%%%
%%%%%%%%%%%%%%%%%%%%%%%%%%
\bibliography{Robustness_particle_creation_ref}

\begin{thebibliography}{11}%
\makeatletter
\providecommand \@ifxundefined [1]{%
 \@ifx{#1\undefined}
}%
\providecommand \@ifnum [1]{%
 \ifnum #1\expandafter \@firstoftwo
 \else \expandafter \@secondoftwo
 \fi
}%
\providecommand \@ifx [1]{%
 \ifx #1\expandafter \@firstoftwo
 \else \expandafter \@secondoftwo
 \fi
}%
\providecommand \natexlab [1]{#1}%
\providecommand \enquote  [1]{``#1''}%
\providecommand \bibnamefont  [1]{#1}%
\providecommand \bibfnamefont [1]{#1}%
\providecommand \citenamefont [1]{#1}%
\providecommand \href@noop [0]{\@secondoftwo}%
\providecommand \href [0]{\begingroup \@sanitize@url \@href}%
\providecommand \@href[1]{\@@startlink{#1}\@@href}%
\providecommand \@@href[1]{\endgroup#1\@@endlink}%
\providecommand \@sanitize@url [0]{\catcode `\\12\catcode `\$12\catcode
  `\&12\catcode `\#12\catcode `\^12\catcode `\_12\catcode `\%12\relax}%
\providecommand \@@startlink[1]{}%
\providecommand \@@endlink[0]{}%
\providecommand \url  [0]{\begingroup\@sanitize@url \@url }%
\providecommand \@url [1]{\endgroup\@href {#1}{\urlprefix }}%
\providecommand \urlprefix  [0]{URL }%
\providecommand \Eprint [0]{\href }%
\providecommand \doibase [0]{http://dx.doi.org/}%
\providecommand \selectlanguage [0]{\@gobble}%
\providecommand \bibinfo  [0]{\@secondoftwo}%
\providecommand \bibfield  [0]{\@secondoftwo}%
\providecommand \translation [1]{[#1]}%
\providecommand \BibitemOpen [0]{}%
\providecommand \bibitemStop [0]{}%
\providecommand \bibitemNoStop [0]{.\EOS\space}%
\providecommand \EOS [0]{\spacefactor3000\relax}%
\providecommand \BibitemShut  [1]{\csname bibitem#1\endcsname}%
\let\auto@bib@innerbib\@empty
%</preamble>
\bibitem [{\citenamefont {Hawking}(1974)}]{hawking1974}%
  \BibitemOpen
  \bibfield  {author} {\bibinfo {author} {\bibfnamefont {S.~W.}\ \bibnamefont
  {Hawking}},\ }\href {\doibase 10.1038/248030a0} {\bibfield  {journal}
  {\bibinfo  {journal} {Nature}\ }\textbf {\bibinfo {volume} {248}},\ \bibinfo
  {pages} {30} (\bibinfo {year} {1974})}\BibitemShut {NoStop}%
\bibitem [{\citenamefont {Hawking}(1975)}]{hawking1975}%
  \BibitemOpen
  \bibfield  {author} {\bibinfo {author} {\bibfnamefont {S.~W.}\ \bibnamefont
  {Hawking}},\ }\href {https://projecteuclid.org:443/euclid.cmp/1103899181}
  {\bibfield  {journal} {\bibinfo  {journal} {Comm. Math. Phys.}\ }\textbf
  {\bibinfo {volume} {43}},\ \bibinfo {pages} {199} (\bibinfo {year}
  {1975})}\BibitemShut {NoStop}%
\bibitem [{\citenamefont {Barcel{\'o}}\ \emph {et~al.}(2011)\citenamefont
  {Barcel{\'o}}, \citenamefont {Liberati}, \citenamefont {Sonego},\ and\
  \citenamefont {Visser}}]{Barcelo2011}%
  \BibitemOpen
  \bibfield  {author} {\bibinfo {author} {\bibfnamefont {C.}~\bibnamefont
  {Barcel{\'o}}}, \bibinfo {author} {\bibfnamefont {S.}~\bibnamefont
  {Liberati}}, \bibinfo {author} {\bibfnamefont {S.}~\bibnamefont {Sonego}}, \
  and\ \bibinfo {author} {\bibfnamefont {M.}~\bibnamefont {Visser}},\ }\href
  {\doibase 10.1007/JHEP02(2011)003} {\bibfield  {journal} {\bibinfo  {journal}
  {Journal of High Energy Physics}\ }\textbf {\bibinfo {volume} {2011}},\
  \bibinfo {pages} {1} (\bibinfo {year} {2011})}\BibitemShut {NoStop}%
\bibitem [{\citenamefont {Cardoso}\ and\ \citenamefont
  {Pani}(2019)}]{cardoso2019testing}%
  \BibitemOpen
  \bibfield  {author} {\bibinfo {author} {\bibfnamefont {V.}~\bibnamefont
  {Cardoso}}\ and\ \bibinfo {author} {\bibfnamefont {P.}~\bibnamefont {Pani}},\
  }\href@noop {} {\enquote {\bibinfo {title} {Testing the nature of dark
  compact objects: a status report},}\ } (\bibinfo {year} {2019}),\ \Eprint
  {http://arxiv.org/abs/1904.05363} {arXiv:1904.05363 [gr-qc]} \BibitemShut
  {NoStop}%
\bibitem [{\citenamefont {Harada}\ \emph {et~al.}(2019)\citenamefont {Harada},
  \citenamefont {Cardoso},\ and\ \citenamefont {Miyata}}]{PhysRevD.99.044039}%
  \BibitemOpen
  \bibfield  {author} {\bibinfo {author} {\bibfnamefont {T.}~\bibnamefont
  {Harada}}, \bibinfo {author} {\bibfnamefont {V.}~\bibnamefont {Cardoso}}, \
  and\ \bibinfo {author} {\bibfnamefont {D.}~\bibnamefont {Miyata}},\ }\href
  {\doibase 10.1103/PhysRevD.99.044039} {\bibfield  {journal} {\bibinfo
  {journal} {Phys. Rev. D}\ }\textbf {\bibinfo {volume} {99}},\ \bibinfo
  {pages} {044039} (\bibinfo {year} {2019})}\BibitemShut {NoStop}%
\bibitem [{\citenamefont {Barcel{\'{o}}}\ \emph {et~al.}(2019)\citenamefont
  {Barcel{\'{o}}}, \citenamefont {Boyanov}, \citenamefont {Carballo-Rubio},\
  and\ \citenamefont {Garay}}]{Barcel__2019}%
  \BibitemOpen
  \bibfield  {author} {\bibinfo {author} {\bibfnamefont {C.}~\bibnamefont
  {Barcel{\'{o}}}}, \bibinfo {author} {\bibfnamefont {V.}~\bibnamefont
  {Boyanov}}, \bibinfo {author} {\bibfnamefont {R.}~\bibnamefont
  {Carballo-Rubio}}, \ and\ \bibinfo {author} {\bibfnamefont {L.~J.}\
  \bibnamefont {Garay}},\ }\href {\doibase 10.1088/1361-6382/ab2e43} {\bibfield
   {journal} {\bibinfo  {journal} {Classical and Quantum Gravity}\ }\textbf
  {\bibinfo {volume} {36}},\ \bibinfo {pages} {165004} (\bibinfo {year}
  {2019})}\BibitemShut {NoStop}%
\bibitem [{\citenamefont {Kokubu}\ and\ \citenamefont
  {Harada}(2019)}]{PhysRevD.100.084028}%
  \BibitemOpen
  \bibfield  {author} {\bibinfo {author} {\bibfnamefont {T.}~\bibnamefont
  {Kokubu}}\ and\ \bibinfo {author} {\bibfnamefont {T.}~\bibnamefont
  {Harada}},\ }\href {\doibase 10.1103/PhysRevD.100.084028} {\bibfield
  {journal} {\bibinfo  {journal} {Phys. Rev. D}\ }\textbf {\bibinfo {volume}
  {100}},\ \bibinfo {pages} {084028} (\bibinfo {year} {2019})}\BibitemShut
  {NoStop}%
\bibitem [{\citenamefont {Ford}\ and\ \citenamefont
  {Parker}(1978)}]{1978_Ford_Parker}%
  \BibitemOpen
  \bibfield  {author} {\bibinfo {author} {\bibfnamefont {L.~H.}\ \bibnamefont
  {Ford}}\ and\ \bibinfo {author} {\bibfnamefont {L.}~\bibnamefont {Parker}},\
  }\href {\doibase 10.1103/PhysRevD.17.1485} {\bibfield  {journal} {\bibinfo
  {journal} {Phys. Rev. D}\ }\textbf {\bibinfo {volume} {17}},\ \bibinfo
  {pages} {1485} (\bibinfo {year} {1978})}\BibitemShut {NoStop}%
\bibitem [{\citenamefont {Birrell}\ and\ \citenamefont
  {Davies}(1982)}]{birrell_davies_1982}%
  \BibitemOpen
  \bibfield  {author} {\bibinfo {author} {\bibfnamefont {N.~D.}\ \bibnamefont
  {Birrell}}\ and\ \bibinfo {author} {\bibfnamefont {P.~C.~W.}\ \bibnamefont
  {Davies}},\ }\href {\doibase 10.1017/CBO9780511622632} {\emph {\bibinfo
  {title} {Quantum Fields in Curved Space}}},\ Cambridge Monographs on
  Mathematical Physics\ (\bibinfo  {publisher} {Cambridge University Press},\
  \bibinfo {year} {1982})\BibitemShut {NoStop}%
\bibitem [{\citenamefont {Hayward}(1996)}]{PhysRevD.53.1938}%
  \BibitemOpen
  \bibfield  {author} {\bibinfo {author} {\bibfnamefont {S.~A.}\ \bibnamefont
  {Hayward}},\ }\href {\doibase 10.1103/PhysRevD.53.1938} {\bibfield  {journal}
  {\bibinfo  {journal} {Phys. Rev. D}\ }\textbf {\bibinfo {volume} {53}},\
  \bibinfo {pages} {1938} (\bibinfo {year} {1996})}\BibitemShut {NoStop}%
\bibitem [{\citenamefont {Oppenheimer}\ and\ \citenamefont
  {Snyder}(1939)}]{PhysRev.56.455}%
  \BibitemOpen
  \bibfield  {author} {\bibinfo {author} {\bibfnamefont {J.~R.}\ \bibnamefont
  {Oppenheimer}}\ and\ \bibinfo {author} {\bibfnamefont {H.}~\bibnamefont
  {Snyder}},\ }\href {\doibase 10.1103/PhysRev.56.455} {\bibfield  {journal}
  {\bibinfo  {journal} {Phys. Rev.}\ }\textbf {\bibinfo {volume} {56}},\
  \bibinfo {pages} {455} (\bibinfo {year} {1939})}\BibitemShut {NoStop}%
\end{thebibliography}%
%\bibliographystyle{JHEP}

%~~\\

%%%%%%%%%%%%%%%%%%%%%%%%%%
%%%%%%%%%%%%%%%%%%%%%%%%%%
%%%%%%%%%%%%%%%%%%%%%%%%%%
%%%%%%%%%%%%%%%%%%%%%%%%%%
%%%%%%%%%%%%%%%%%%%%%%%%%%
%%%%%%%%%%%%%%%%%%%%%%%%%%
%%%%%%%%%%%%%%%%%%%%%%%%%%
%%%%%%%%%%%%%%%%%%%%%%%%%%
\end{document}